\documentclass[a4paper,11pt]{article}
\usepackage{pos}
\usepackage{xcolor}

\title{Machine-learning approaches to accelerating lattice simulations}

\author*{Scott Lawrence}

\affiliation{Los Alamos National Laboratory Theoretical Division T-2,\\Los Alamos, NM 87545, USA}

\emailAdd{srlawrence@lanl.gov}

\abstract{The last decade has seen an explosive growth of interest in exploiting developments in machine learning to accelerate lattice QCD calculations. On the sampling side, generative models are a promising approach to mitigating critical slowing down and topological freezing. Meanwhile, signal-to-noise problems have been shown to be improvable by the use of optimized improved observables. Both techniques can be made free of bias, resulting in trustworthy but reduced statistical errors. This talk reviews recent developments in this field.}

\FullConference{The 41st International Symposium on Lattice Field Theory (LATTICE2024)\\
 28 July - 3 August 2024\\
Liverpool, UK\\}

\tableofcontents

\begin{document}
\maketitle

\section{Introduction}
Lattice Monte Carlo methods (particularly lattice QCD) allow nonperturbative, first-principles access to a wide range of observables in strongly coupled systems, but at considerable computational cost. Algorithmic improvements that reduce this cost, while preserving the merits of these Monte Carlo methods---particularly controllable systematic errors, predictable polynomial scaling, and access to nonperturbative physics---are valuable. Recent searches for such improvements have often been inspired by techniques from machine learning (ML) and particularly deep learning. To slightly over-generalize, one may describe an algorithmic improvement that depends on the existence of some function $f(x)$ such that a functional $L[f(x)] \approx 0$ is small. The function $f(x)$ is then parameterized and sought via ML techniques (typically an appropriate equivalent of stochastic gradient descent on the loss functional $L[\cdot]$). This approach has the virtue that the last decade's substantial progress in deep learning techniques can be brought to bear on lattice QCD with relatively little modification.

This talk is focused on those ML and ML-inspired methods which accelerate lattice simulations in a way that provably introduces no systematic bias, without significantly altering the overall framework of lattice QCD. The ML-inspired methods discussed in this talk can be divided into two rough families, depending on whether the method is intended to assist with sampling or with measurement. On the sampling side, there have been an array of proposals for learning optimized proposal distributions. The greatest successes in gauge theories have been seen by flow-based samplers, discussed in Section~\ref{sec:flow}. On the measurement side, signal-to-noise problems (and, closely related, sign problems) can be alleviated by performing contour deformations as discussed in Section~\ref{sec:contours}, or by explicitly constructing observables with equal mean but reduced variance as discussed in Section~\ref{sec:cv}. Finally, measurement can be made cheaper without great sacrifices to precision by learning approximate observables which may be quickly evaluated, while using a small number of configurations to remove the bias (see Section~\ref{sec:surrogates}).

Of necessity a wide variety of techniques have been excluded from this talk, not because of a lack of merit but because they do not fit with the narrow framework described above. To conclude this section let us highlight a portion of this work. Various machine learning methods have been applied to regularize the ill-posed problem of determining spectral functions~\cite{Chen:2021giw,SamuelOffler:2022jkv,PhysRevLett.124.056401,PhysRevD.102.096001,Wang:2021jou}, relevant to determining real-time evolution, decay rates, and cross sections from Euclidean lattice data. In something of a call-back to the origins of deep learning of a field, phases or action parameters can be learned from samples (or a single sample) of field configurations~\cite{carrasquilla2017machine,Peng:2022wdl,Tanaka:2016rtu,vanNieuwenburg:2016zsd,Rodriguez-Nieva:2018cbl,Broecker:2017hjl,Shanahan:2018vcv}. Following the introduction of neural-network quantum states~\cite{carleo2017solving}, neural networks have been used to represent wavefunctions of lattice systems (in the Hamiltonian formalism)~\cite{Deng:2017wof,Luo:2020stn}.
		Finally, neural networks have recently been used to parameterize a many-parameter lattice action~\cite{Holland:2024muu} with an eye towards minimizing lattice discretization effects.

\section{Flow-based sampling}\label{sec:flow}

A substantial fraction of the computational effort of lattice QCD comes from the cost of obtaining a sequences of $N$ samples $U_n$ from the Boltzmann distribution $p(U) \propto e^{-S_{\mathrm{eff}}(U)}$. The statistical errors on a Monte Carlo measurement scale as $\sigma \sim N^{-1/2}$, so assuming negligible systematics and cheap measurement\footnote{Neither is a universally valid assumption: many calculations have errors dominated by systematics, and measurement is typically at least comparably expensive to sampling due to the need to invert the Dirac matrix.}, a method that accelerates sampling by a factor $A$ will reduce error bars by a factor $\sqrt A$.

Thus we are motivated to consider the vast family of machine-learning methods for efficiently sampling from various probability distributions. The focus here will be on one particular class of such methods: normalizing flows (also highlighted in a plenary talk at the previous lattice conference~\cite{Kanwar:2024ujc}). Several other ML methods for accelerating sampling have been explored in recent years, including restricted Boltzmann machines~\cite{2017PhRvB..95c5105H,Tanaka:2017niz}, diffusion models~\cite{Wang:2023exq,Wang:2023sry}, and generative adversarial networks~\cite{Zhou:2018ill,Pawlowski:2018qxs}. Our focus is on normalizing flows because normalizing flows have been tested directly on gauge theories---the performance of other methods on gauge models of size comparable to QCD lattice is generally unexplored.

Let us say we have a probability distribution on $\mathbb R$, $p(x)$, from which we wish to sample. A normalizing flow for this distribution is a function $x(z)$ that obeys:
\begin{equation}
	p(x) dx = \frac{1}{\sqrt {2\pi}} e^{-z^2/2} dz
	\text.
\end{equation}
A normally distributed random variable $z$ can be turned into a sample from $p$ simply by applying the function $x(z)$. Normalizing flows exist for any distribution $p$, and can be obtained in one dimension by inverting the cumulative distribution function.

Normalizing flows naturally generalize to multivariate distributions, where the defining equation is
\begin{equation}\label{eq:nf-definition}
	p[\phi(z)] \det \frac{\partial\phi}{\partial z} = \frac{1}{\sqrt{2\pi}} e^{-z^2 / 2}
	\text.
\end{equation}
An early published example of a normalizing flow is the Box-M\"uller transform~\cite{box1958note}, which converts between a uniform sample on $[0,1]^2$ and two normally distributed random variables.
\begin{figure}
	\centering
	\includegraphics[width=0.9\linewidth]{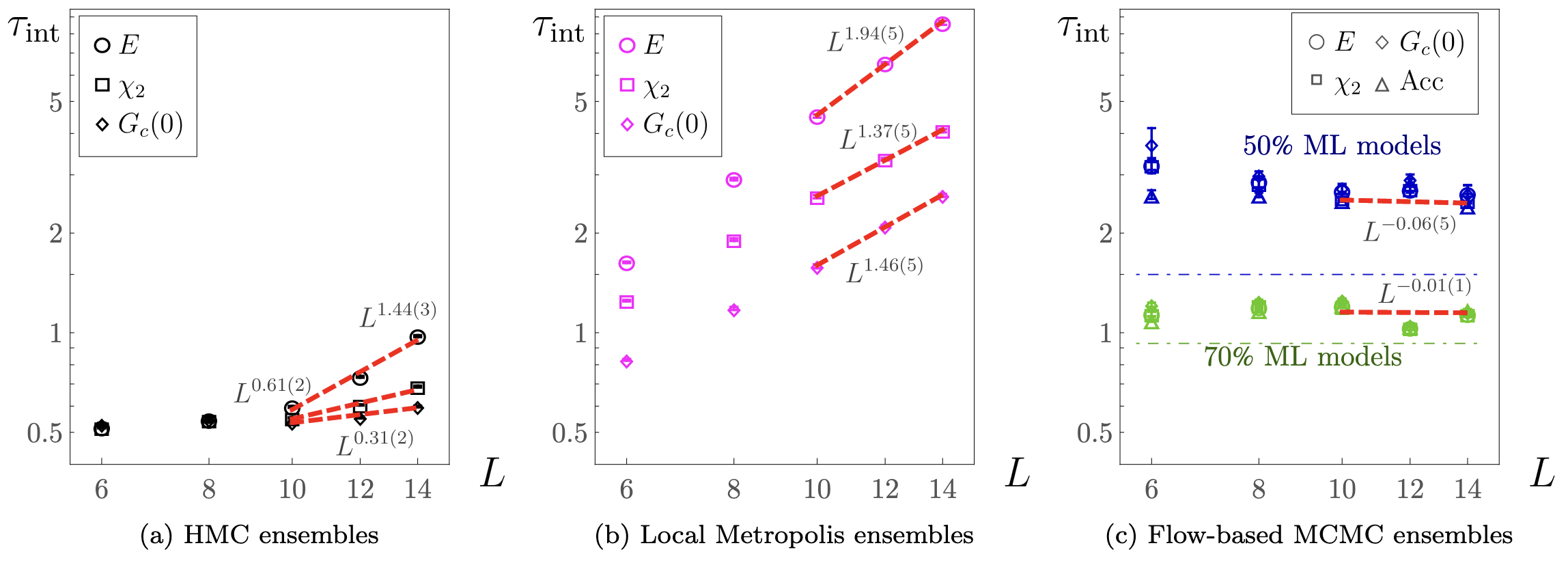}
	\caption{Comparison of the autocorrelation time as a function of lattice size when using HMC updates, local Metropolis updates, and normalizing flows. Reproduced from~\cite{Albergo:2019eim}.\label{fig:flow-autocorrelation}}
\end{figure}
In physics, normalizing flows first appeared under the name of \emph{trivializing maps} as a method for solving field theories. Nicolai presented one as a solution to $\mathcal N=4$ supersymmetric Yang-Mills~\cite{Nicolai:1979nr}, and three decades later Luscher discussed trivializing maps in the context of lattice gauge theory~\cite{Luscher:2009eq}.

So far all of these normalizing flows have been \emph{exact} normalizing flows, meaning that (\ref{eq:nf-definition}) holds as written. When attempting to learn a normalizing flow, one does not typically expect to be able to satisfy such an equality exactly. We instead search for flows $\phi(z)$ such that (\ref{eq:nf-definition}) is satisfied only approximately. Expectation values over the original distribution $p(\cdot)$ are then estimated either by using the normalizing flow to generate proposals in a Markov-chain Monte Carlo, or by reweighting.
For good approximations, this is often still much more efficient than sampling from $p(\cdot)$ by other means.

Training and using a numerical normalizing flow involves repeated evaluations of the Jacobian determinant. For an arbitrary function $\phi$ this calculation scales nearly as the cube of the number of degrees of freedom, and is by far the most computationally intensive step. 
	Therefore, practical normalizing flows use a restricted class of functions for which both $\phi$ and its Jacobian determinant can be evaluated efficiently---typically in time linear in the number of degrees of freedom. These families include NICE~\cite{2014arXiv1410.8516D}, RealNVP~\cite{2016arXiv160508803D}, and neural ordinary differential equations~\cite{2018arXiv180703039K}.

		Continuous normalizing flows have been used to sample the Nambu-Goto string~\cite{Caselle:2023mvh} and to construct normalizing flows the exactly respect the symmetries of the field theory~\cite{deHaan:2021erb}. However, most large-scale numerical works have used RealNVP.

		A trained normalizing flow allows the partition function to be directly estimated~\cite{Nicoli:2019gun,Nicoli:2020njz}. In the context of lattice quantum field theory, the partition function is typically not of physical interest, and so this property is not much used.

		The use of RealNVP (and various extensions) has led to a series of remarkable results in accelerating lattice sampling. The first demonstration of this was in scalar field theory~\cite{Albergo:2019eim}. In that work, the flow is used to generate proposals for a Markov-chain Monte Carlo ensemble, and so the success of the normalizing flow manifests as a short autocorrelation time. This is shown in Figure~\ref{fig:flow-autocorrelation}.

		Applications to gauge theories followed from the design of gauge-equivariant flows~\cite{Boyda:2020hsi,Kanwar:2020xzo,Kanwar:2021wzm}, which are only capable of representing gauge-invariant probability distributions~\footnote{Note that this is not because gauge-equivariance is required for correctness, but rather because it is believed to be easier to train a flow which already ``knows about'' all relevant symmetries.}. This has since been extened to theories with fermions~\cite{Abbott:2022zhs}, applied to the lattice Schwinger model~\cite{Albergo:2022qfi}, and demonstrated at arbitrary space-time dimension~\cite{Abbott:2023thq}.
	\begin{table}
		\centering
		\begin{tabular}{ccc}
			Reference & $N_\mathrm{params}$ & ESS at $\beta = 6$\\\hline
			\cite{Luscher:2009eq} & 8 & $<1\%$ \\
			\cite{Bacchio:2022vje} & 420 & $70\%$ \\
			\cite{Boyda:2020hsi} & $\sim 10^6$ & $48\%$
		\end{tabular}
		\caption{Comparison of the performance of three different normalizing flows, as measured by effective sample size (ESS). Larger ESS corresponds to a higher-precision normalizing flow and a more efficient algorithm. Table reproduced from~\cite{Bacchio:2022vje}\label{tab:flows}}
	\end{table}

	Most works in this line have adopted the usual perspective of deep learning, that more parameters leads to better results. At least in the case of $1+1$-dimensional gauge theory (a special case, being an exactly solvable model), this is not necessarily true. In~\cite{Bacchio:2022vje}, it was demonstrated that a few-parameter ansatz (inspired by L\"uscher's flows) is capable of substantially out-performing numerically trained flows with $\sim 10^6$ parameters. The key table from this work is reproduced as Table~\ref{tab:flows}.

\section{Contour deformations}\label{sec:contours}
We now turn from methods designed to accelerate sampling, to those designed to modify the signal-to-noise ratio after sampling has been completed. The method of \emph{contour deformations} is based on Cauchy's integral theorem. In the case of a single complex dimension, for any contour $\gamma$ obtained by continuously deforming $\mathbb R \subset \mathbb C$, we have the equality
\begin{equation}
	\int_{\mathbb R} f(z) dz = \int_{\gamma} f(z) dz
\end{equation}
for any function $f(z)$ which is holomorphic. The complex integration measure is defined as $dz = dx + i dy$. The two integrals are equal, but the values the integrand takes along $\mathbb R$ are not the same as those the integrand takes along the contour $\gamma$. As a result, one integral may be numerically easier than the other.

In the $N$-dimensional case, Cauchy's integral theorem states\footnote{As pointed out in~\cite{Lawrence:2022afv}, this construction excludes those contours homologous but not homotopic to the real plane. No such contour has yet been used to mitigate a sign problem.}
\begin{equation}
	\int_{\mathbb R^N} f(z) dz = \int_{\mathbb R^N} f(\phi(x)) \det \frac{\partial \phi}{\partial x} dz
\end{equation}
for any continuous (and piecewise differentiable) $\phi : \mathbb R^N \rightarrow \mathbb C^N$ parameterizing an $N$-dimensional integration contour. Suitable generalizations to functions defined on manifolds that only locally look like $\mathbb R^N$ are readily available. This includes the integrals over $SU(N)$ which are relevant to lattice QCD.

This fact has been applied to both sign problems (where the Boltzmann factor itself posesses strong fluctuations) and signal-to-noise problems, in slightly different ways.

\subsection{Sign problems}
\begin{figure}
	\centering
	\includegraphics[width=5in]{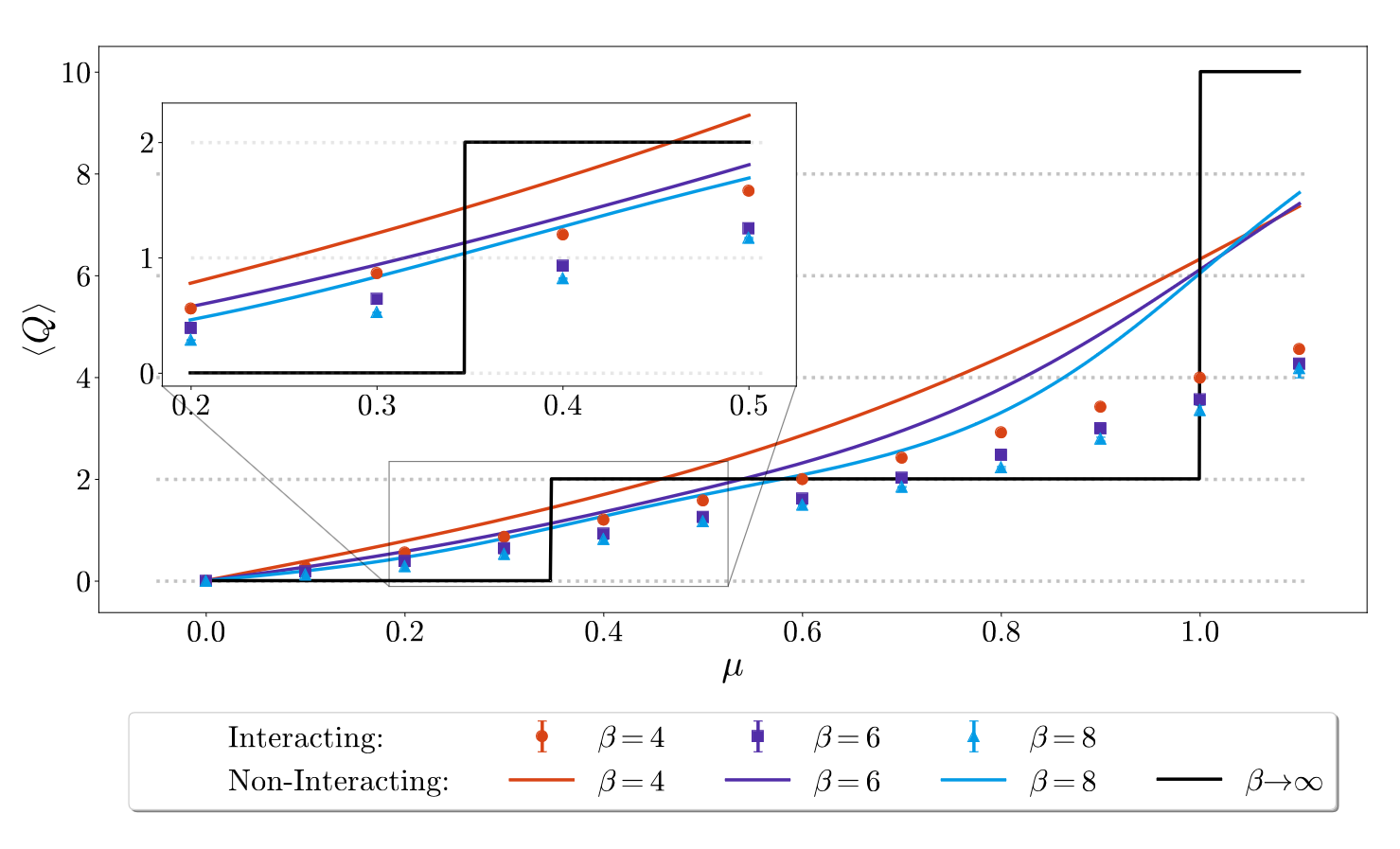}
	\caption{Reproduced from~\cite{Rodekamp:2024ixu}, the charge of Perylene as a function of chemical potential. Computed with a contour deformation, defined by one parameter.\label{fig:perylene}}
\end{figure}
Historically, the motivating applications of contour deformations were theories that exhibit \emph{sign problems}.  In short\footnote{See~\cite{Alexandru:2020wrj} for a review.}, we have a theory for which the action is in general complex, and therefore the Boltzmann factor $e^{-S}$ no longer defines a probability distribution. In this context expectation values are commonly computed by sampling with respect to the ``quenched ensemble'', defined by a probability distribution proportional to $|e^{-S}|$. This comes at a severe cost. Defining the \emph{average phase}
\begin{equation}\label{eq:average-phase}
	\langle \sigma \rangle
	\equiv
	\frac{\int e^{-S(U)} dU}{\int |e^{-S(U)}|dU}
	\equiv \frac{Z}{Z_Q}\text,
\end{equation}
the number of samples required to overcome the sign problem scales as $\langle \sigma \rangle^{-2}$. As the average phase typically decays exponentially with the system volume, this imposes an exponential cost on lattice calculations.

Now consider performing a contour deformation---instead of integrating over $\mathbb SU(3)^{\otimes V}$, we integrate over some contour $\gamma \subset SL(3;\mathbb C)^{\otimes V}$. Expectation values are expressible as a ratio of two holomorphic integrals, and are therefore not affected by contour deformation:
\begin{equation}
	\langle \mathcal O \rangle =
	\frac{\int_{SU(3)^{\otimes V}} \mathcal O(U) e^{-S(U)} d^VU}{\int_{SU(3)^{\otimes V}}e^{-S(U)} d^VU}
	=
	\frac{\int_{\gamma} \mathcal O(M) e^{-S(M)} d^VM}{\int_{\gamma}e^{-S(M)} d^VM}
	\text.
\end{equation}
The figure of merit above, however---the average phase---cannot be written in this way. The quenched partition function $Z_Q$ is dependent on the choice of contour. We therefore conclude that it is possible to improve the sign problem by judicious choice of integration contour.

It remains to find a useful integration contour: one on which $\langle \sigma\rangle$ is as large as possible. Inspired by early work on analytic continuation of field theories~\cite{Witten:2010cx}, the first proposed contours were Lefschetz thimbles~\cite{Cristoforetti:2012su}: a particular family of contours for which the imaginary part of the action happens to be constant\footnote{To prevent a common misconception, note that the Lefschetz thimbles do not in general maximize $\langle \sigma\rangle$. This is both because the Jacobian determinant introduces additional phase fluctuations, and because of cancellations between the thimbles.}. The Lefschetz thimbles are computationally expensive to find, and so it was proposed to use supervised learning to obtain an efficient approximation~\cite{Alexandru:2017czx}.

Shortly after, it was realized that instead of attempting to approximate the Lefschetz thimbles, one could directly maximize the average phase $\langle \sigma\rangle$ over a family of contour deformations~\cite{Mori:2017pne,Alexandru:2018fqp}.
This unsupervised learning algorithm has been applied to a variety of systems, including fermions at finite density in $2+1$ dimensions~\cite{Alexandru:2018ddf}. 

One lesson learned from the many works searching for integration contours with a mild sign problem, is that very simple contour deformations are often sufficient in practice. The simplest nontrivial contour deformation is a ``constant shift'', where each degree of freedom has the same imaginary part added:
\[
	\phi_x \rightarrow \phi_x + i C
	\text.
\]
This sort of contour deformation has proven useful in the study of small systems of strongly correlated electrons~\cite{Gantgen:2023byf}. Note that when performing a Monte Carlo on a deformed contour, there is a Jacobian determinant that must be evaluated, coming from the curvature of the contour. A constant shift has no curvature, and so there is no cost associated to this determinant. This line of work culminated in the calculation of the behavior and spectrum of doped Perylene~\cite{Rodekamp:2024ixu}. From that work, the charge as a function of chemical potential is shown in Figure~\ref{fig:perylene}.

\begin{figure}
	\centering
	\includegraphics[width=0.48\linewidth]{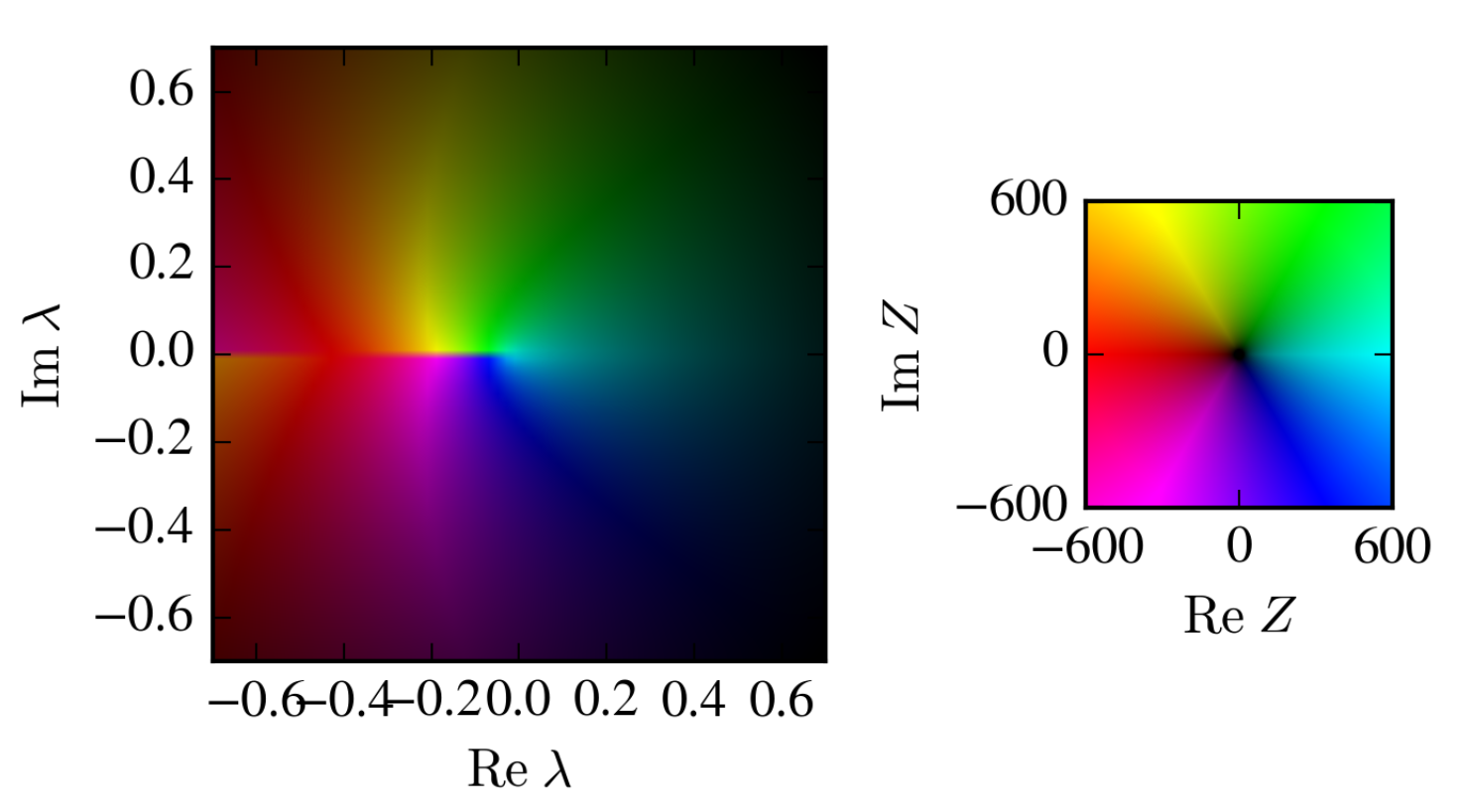}
	\hfill
	\includegraphics[width=0.48\linewidth]{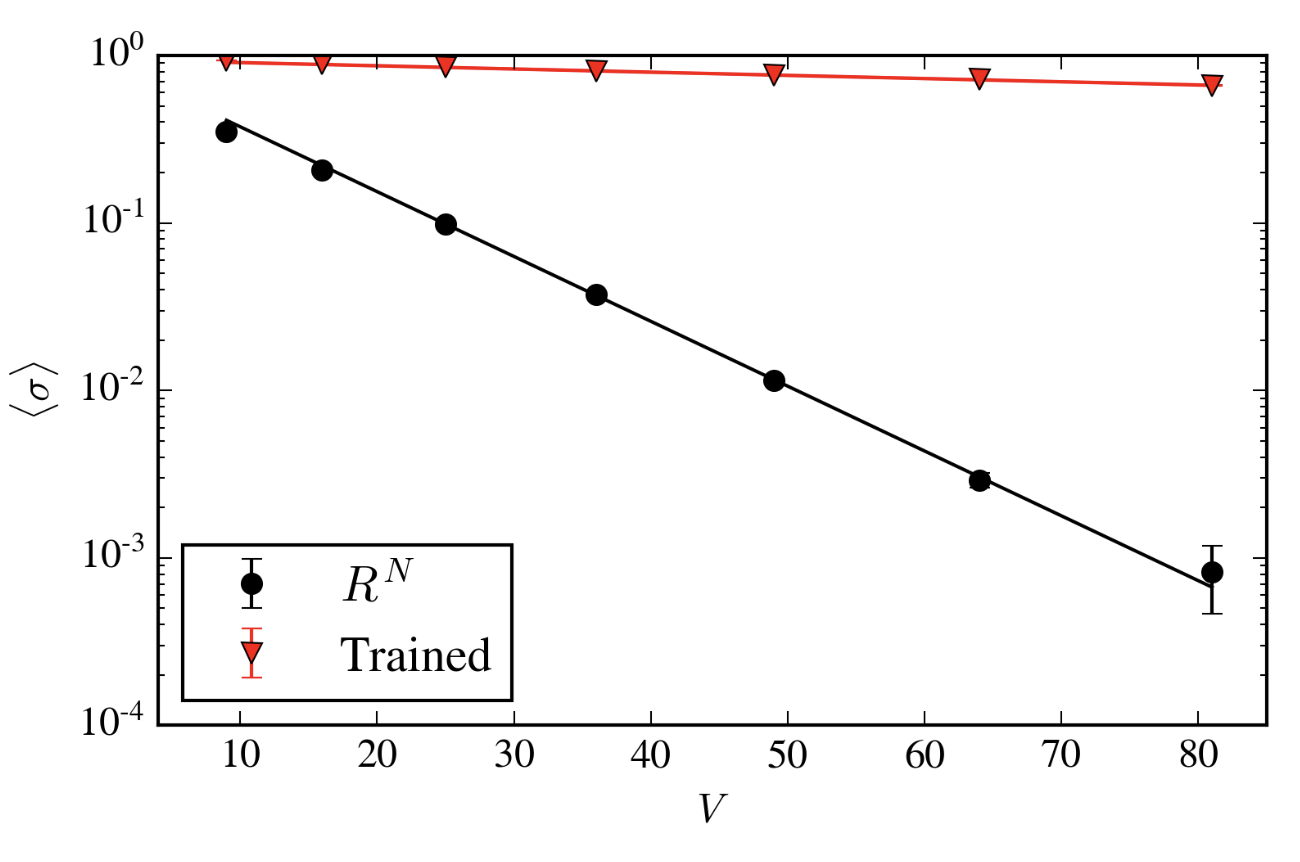}
	\caption{From~\cite{Lawrence:2022afv}, investigations of lattice scalar field theory at complex coupling. At left is shown the partition function of $0+1$-dimensional scalar ($\lambda \phi^4$) field theory as a function of complex anharmonicity $\lambda$; the branch cut on $\mathbb R_-$ can clearly be seen. At right, the average phase of $1+1$-dimensional lattice scalar field theory at $\lambda=i$, as a function of volume. Both the defining contour $\mathbb R^V$ and the trained contour exhibit exponential decays, but the trained contour has a slope more than an order of magnitude smaller.\label{fig:complex-scalar}}
\end{figure}

More complicated contours, defined by many-parameter neural networks, were demonstrated to be valuable for the calculation of scalar $\lambda \phi^4$ field theory at complex coupling~\cite{Lawrence:2022afv}.
This model is of possible interest as a $\mathcal {PT}$-symmetric field theory~\cite{Bender:2018pbv,Romatschke:2024cld,Lawrence:2023woz,Weller:2023jhc}.
Key figures from this work are reproduced in Figure~\ref{fig:complex-scalar}.

These sorts of machine-learned contour deformations have also been used to study the $XY$ model in $2+1$ dimensions~\cite{Tulipant:2022vtk}, $U(1)$ gauge theory with complex parameters~\cite{Kashiwa:2020brj}, one-dimensional QCD~\cite{Basar:2022cef}, QCD in the heavy-dense limit~\cite{Basar:2023bwd}.

As with normalizing flows (discussed above), it is generally believed that an ansatz will perform better if the symmetries of the problem are ``built-in'', so that they do not have to be learned during training. This has provided motivation for the study of the performance of neural networks that take as input only gauge-invariant quantities~\cite{Namekawa:2021nzu}.

Not all successes of the contour deformation method have stemmed from machine learning. A popular alternative is the \emph{holomorphic gradient flow}, which defines a family of contours that approximate the Lefschetz thimbles. This family has been used to study the chiral random matrix model~\cite{Giordano:2023ppk} and QED in $1+1$ dimensions~\cite{Alexandru:2018ngw}.

The contour deformation method, on its face, is only able to handle systems with continuous parameters (which can then be complexified). This can be extended to spin systems by introducing auxiliary degrees of freedom~\cite{Kashiwa:2023dfx,Warrington:2023aqa,Mooney:2021esz}.

Finally, most of the works above have focused on systems at a finite chemical potential (which introduces the sign problem). Via the lattice Schwinger-Keldysh formalism, real-time dynamics can also be studied, albeit with a severe sign problem. This sign problem has been the target of various contour deformation-based methods~\cite{Alexandru:2017lqr,Alexandru:2016gsd,Warrington:2023aqa}, including machine learning~\cite{Lawrence:2021izu}. However, in the case of field variables on compact spaces (particularly relevant to gauge theory), there are substantial technical obstacles to preserving unitarity in time-evolution~\cite{Kanwar:2021tkd}.

\subsection{Signal-to-noise problems}
Sign problems and signal-to-noise problems are closely related. In fact, a sign problem can be seen as a signal-to-noise problem in evaluating the ratio of partition functions $Z / Z_Q$. Just as a sign problem can be improved by choice of appropriate contour (because $Z_Q$ is not the integral of a holomorphic quantity), so can the noise in an observable be reduced by choice of an appropriate integration contour. This comes from the fact that the variance associated to $\langle \mathcal O \rangle$ is $\langle \mathcal O^\dagger \mathcal O \rangle$---the operator in the expectation value is not holomorphic.

\begin{figure}
	\centering
	\includegraphics[width=3in]{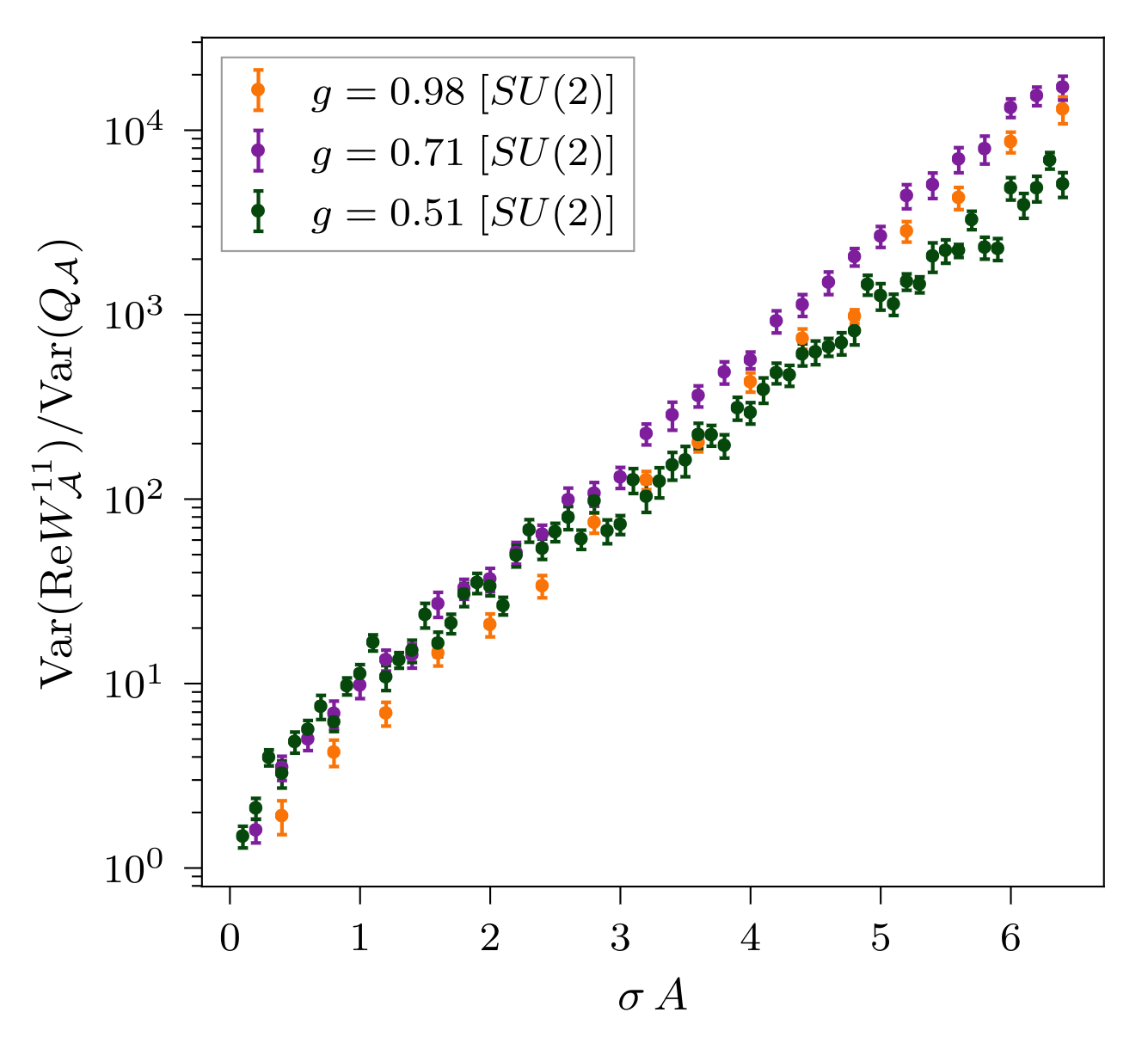}
	\caption{From~\cite{Detmold:2021ulb}, evidence of exponential improvement in the signal-to-noise problem of two-dimensional $SU(2)$ Yang-Mills. Shown is the ratio of the variance of the estimator of the Wilson loop, to the the variance of a contour-improved estimator of the same quantity.\label{fig:stn}}
\end{figure}

Thus motivated, simple contour deformations have been explored in the context of $U(1)$ gauge theory and complex scalar field theory~\cite{Detmold:2020ncp}, and two-dimensional $SU(2)$ Yang-Mills~\cite{Detmold:2021ulb}. A demonstration from the latter work is reproduced as Figure~\ref{fig:stn}. The improvement in the signal-to-noise ratio is shown to be exponential in the size of the Wilson loop.

\subsection{Theoretical aspects}
Beyond the machine-learning aspects of this method, a great deal is now known about the space of possible contour deformations and their sign problems. Here we briefly summarize a few important results and major open questions.

If, in the definition (\ref{eq:nf-definition}) of a normalizing flow, the function $x : \mathbb R^N \rightarrow \mathbb C^N$ is permitted to be complex-valued, then $x(z)$ can be seen to parameterize a complex integration contour. This observation leads to the notion of a \emph{complex normalizing flow}~\cite{Lawrence:2021izu}, which can be used to work around the difficulty of efficiently sampling on an arbitrary contour deformation. In the context of machine learning, by learning a complex normalizing flow, one obtains many of the benefits of a contour deformation and a normalizing flow simultaneously.

The discussion of contour deformations above hinged on the Boltzmann factor $e^{-S}$ being holomorphic. For some applications (e.g.~nuclear forces) this assumption does not hold. It has been shown that this approximation can be relaxed, either by approximating the action by a nearby holomorphic one~\cite{Kanwar:2023otc}, or by constructing a Riemann surface on which the Boltzmann factor is holomorphic~\cite{Lawrence:2024pjg}. In both cases, a key insight is that the calculation we want to perform is defined only on the real plane. The behavior of the action (and other functions) away from the real plane is a choice we make for algorithmic convenience.

It has been shown that a necessary and sufficient condition for a contour $\gamma$ to be a \emph{local} maximum of $\langle \sigma\rangle$, is that the integrand $e^{-S} dz$ has locally constant phase as long as that phase is defined. In other words, the phase of $e^{-S} dz$ is piecewise constant, with regions of different phase being separated by zeros of that integrand. This is reminiscent of a similar property of Lefschetz thimbles, where it is the phase of $e^{-S}$ that is constant except where that Boltzmann factor vanishes. It has been conjectured~\cite{Gantgen:2023byf,Lawrence:2023sfc} that there exists an expansion in $\hbar$, of which the Lefschetz thimbles are the leading-order term and the perfect contours are the sum. Rigorous results to this effect are not yet available.

Empirically, most studies that use contour deformations have failed to find a perfect contour. That is, there is usually some residual average phase, which decays exponentially with the volume of the system. When this happens, it is \emph{a priori} unclear whether no such contour exists, or the learning algorithm has simply failed to find the perfect contour (in which case applying more computational power may be advisable). In~\cite{Lawrence:2023sfc}, a method of proving that no perfect contour exists was devised. Writing the quenched partition function over a contour $\gamma$ as $\int_\gamma |\omega|$ for some differential form $\omega$, one finds another differential form $\alpha$ obeying both $d\alpha = 0$ and $|\alpha| \le |\omega|$. For any such form $\alpha$ one has the following sequence of inequalities:
\begin{equation}
	\left|\int_{\mathbb R} \alpha\right| = \left|\int_\gamma \alpha\right| \le \int_{\gamma} |\alpha| \le \int_{\gamma} |\omega|\text,
\end{equation}
establishing that $\int_{\mathbb R}\alpha$ is a lower bound on the quenched partition function, no matter the choice of contour.
\begin{figure}
	\centering
	\includegraphics[width=3in]{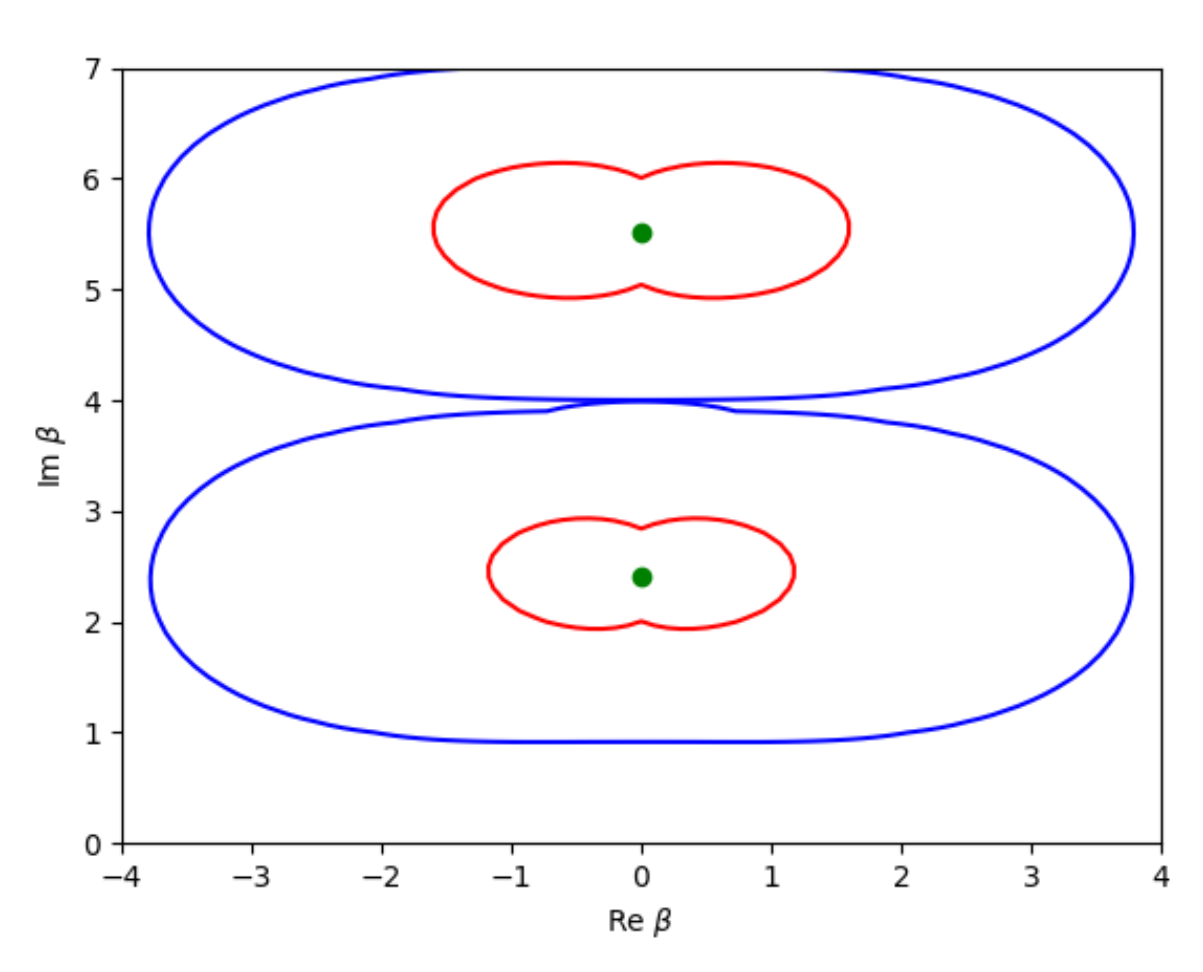}
	\caption{No-go regions for $1+1$-dimensional lattice Yang-Mills at complex coupling $\beta$, from~\cite{Lawrence:2023sfc}. Outside of the blue lines, perfect integration contours (on which $\langle\sigma\rangle = 1$) were found numerically; within the red lines, proofs that no such contours existed were found. The black points indicate zeros of the partition function.\label{fig:nogo}}
\end{figure}
By this method an exponentially decaying upper bound on $\langle \sigma\rangle$ was found for $1+1$-dimensional lattice Yang-Mills for a range of complex couplings (see Figure~\ref{fig:nogo}). No such bounds have yet been found for theories that are not exactly solvable; nor have any such bounds been derived for improvements in signal-to-noise problems.

\section{Control variates}\label{sec:cv}
For the remainder of this talk we continue with the theme of variance reduction. Of course this can often be accomplished via normalizing flows~\cite{Abbott:2024kfc}: the variance is reduced if the number of samples can be efficiently increased. However it may be that no high-quality normalizing flow can be constructed, or that the cost of measurement outweighs the cost of sampling. In such cases we desired to modify the observable in such a way that the variance changes while the expectation value remains the same.

Suppose we know of a function $f$ such that $\langle f \rangle = 0$ vanishes exactly. Then instead of computing $\langle \mathcal O \rangle$, we might computed $\langle \mathcal O - f\rangle$. The two expectation values are exactly equal, but the statistical noise might be improved. The function $f$ is termed a \emph{control variate}. Control variates are extraordinarily general: for every observable there exists a ``perfect'' control variate that entirely removes the noise~\cite{Lawrence:2022dba}. Moreover, other variance reduction methods (including contour deformations as discussed above) can often be re-written as control variates~\cite{Lawrence:2022dba}. This generality is both a blessing and a curse: we know that there exist desirable control variates, but we have little guidance in how to find them.

Control variates are in principle applicable to both sign problems and signal-to-noise problems. The most striking successes of control variates have all been in their application to signal-to-noise problems, so this section will focus on those cases. Small but significant improvements have been obtained in the sign problems for fermionic systems~\cite{Lawrence:2020kyw,Lawrence:2022dba} and Ising models at complex coupling~\cite{Lawrence:2023cft}.

To make this approach practical, we first need a large family of functions $f$ whose expectation values can be proven\footnote{It is critical that $\langle f \rangle = 0$ be a theorem, rather than an empirical statement from Monte Carlo data. By construction, the observable $f$ will be noisy.} to vanish exactly. Such a family of functions is provided by noticing that the integral of a total derivative (on a compact space of field configurations) must vanish. As a result we have
\begin{equation}
	\langle \partial g \rangle = \langle g \partial S \rangle\text,
\end{equation}
where $S$ is the action and $g$ is any function of field configurations. Equalities of this form, which hold for any first-order derivative $\partial$, are termed \emph{Schwinger-Dyson relations}.

The immediate problem with applying this method to lattices of realistic sizes is that there are simply too many Schwinger-Dyson relations, and in a sense we do not know which ones to use. To be concrete, consider lattice scalar field theory with $V$ sites. At leading order there are $V^2$ Schwinger-Dyson relations available, of the form
\begin{equation}
	\langle \frac{\partial}{\partial \phi_x} \phi_y \rangle = \langle \phi_y \frac{\partial}{\partial \phi_x} S\rangle
\end{equation}
for (not necessarily distinct) sites $(x,y)$. Using translational invariance this is reduced to a set of $V$ relations. A general control variate may be written as a linear combination of these. This control variate has $V$ free coefficients, which must be determined by a fit. On a $2+1$-dimensional lattice of reasonable ($\sim 100^3$) size, this fit requires $\gtrsim 10^6$ configurations, which are likely not available.

\begin{figure}
	\centering
	\includegraphics[width=0.48\linewidth]{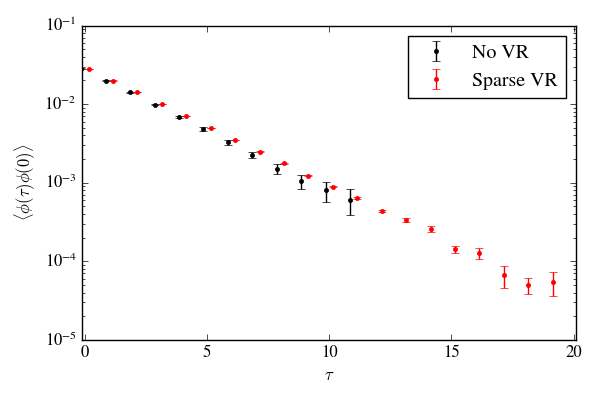}
	\hfill
	\includegraphics[width=0.48\linewidth]{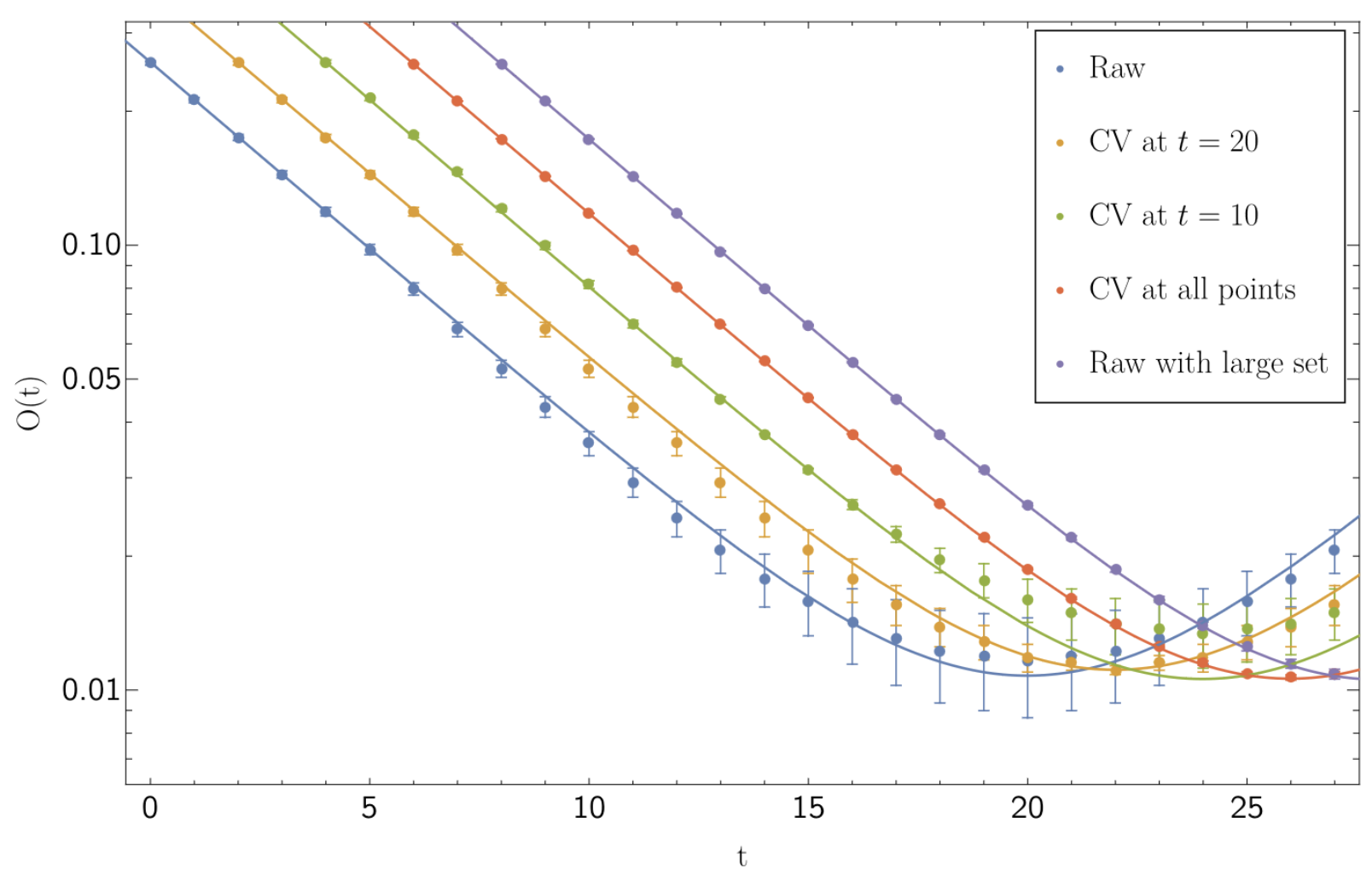}
	\caption{Two approaches to reducing the signal-to-noise problem associated with the correlator in scalar field theory, both based on Schwinger-Dyson control variates. At left~\cite{Bhattacharya:2023pxx}, a complete basis of leading-order Schwinger-Dyson relations is used, with $L_1$ regularization to mitigate overfitting. At right~\cite{Bedaque:2023ovz}, a neural network parameterizes the function $g(\phi)$ which defines the Schwinger-Dyson control variate. \label{fig:cv}}
\end{figure}

One approach, proposed in~\cite{Bhattacharya:2023pxx}, is to use $L_1$ regularization. The assumption is made that a good control variate exists for which the vast majority of coefficients are exactly $0$, and such a ``sparse'' control variate is optimized numerically. The performance of this method is shown in the left panel of Figure~\ref{fig:cv}: an order-of-magnitude improvement in the signal-to-noise ratio, corresponding to a speed-up of a factor of $\sim 10^2$, is obtained.

Another proposed approach to this high-dimensional fit has been to parameterize the function $g$ by a neural network~\cite{Bedaque:2023ovz}. Recall that any function $g$ defines a control variate. The high-dimensional fitting problem can then be approached in any number of ways standard in machine learning, including regularization by limiting training time. Tested in scalar field theory, this method also yields an improvement in the signal-to-noise ratio of more than an order of magnitude (see again Figure~\ref{fig:cv}).

Recent work has also proposed constructing control variates in perturbation theory~\cite{Lawrence:2024xsi}, although with less success than the previously mentioned approaches.

\section{Surrogate observables}\label{sec:surrogates}
The previous section focused on reducing the noise associated with measuring an expectation value $\langle \mathcal O \rangle$. This is valuable when, for any reason, we cannot perform a sufficiently large number of calculations of $\mathcal O(U)$ on independent configurations. This may be because collecting the configurations is expensive (the cost of HMC), or because given $U$, the task of evaluating $\mathcal O(U)$ is expensive. In either case, reducing the noise will reduce the number of configurations on which $\mathcal O(U)$ is needed.

When $\mathcal O(U)$ is expensive and collecting configurations $U$ is (relatively) cheap, another approach is available. By one method or another, we construct an approximation $\tilde{\mathcal O}(U) \approx \mathcal O(U)$, which is cheaper to evaluate. We can evaluate $\tilde{\mathcal O}$ on a larger number of configurations; however its use introduces a systematic bias from the difference $\langle {\mathcal O} \rangle - \langle \tilde{\mathcal O} \rangle \ne 0$. This bias can be corrected by evaluating $\langle \mathcal O - \tilde{\mathcal O}\rangle$. Per-sample, this is as expensive as evaluating $\mathcal O$; however, because this last expectation value has smaller variance, it does not need to be evaluated on as many samples.

This approach is summarized by the equation
\begin{equation}
	\langle \mathcal O\rangle
	=
	\langle \tilde{\mathcal O}\rangle + \langle \mathcal O - \tilde{\mathcal O}\rangle
	\approx
	\langle \tilde{\mathcal O}\rangle_{\tilde N} + \langle \mathcal O - \tilde{\mathcal O}\rangle_N
	\text,
\end{equation}
where $\langle\cdot\rangle$ denotes the true lattice expectation value, and $\langle\cdot\rangle_n$ an expectation value with respect to $n$ samples. The ratio $\frac N {\tilde N} < 1$ controls which term in the estimator dominates the variance.

\begin{figure}
	\centering
	\includegraphics[width=3in]{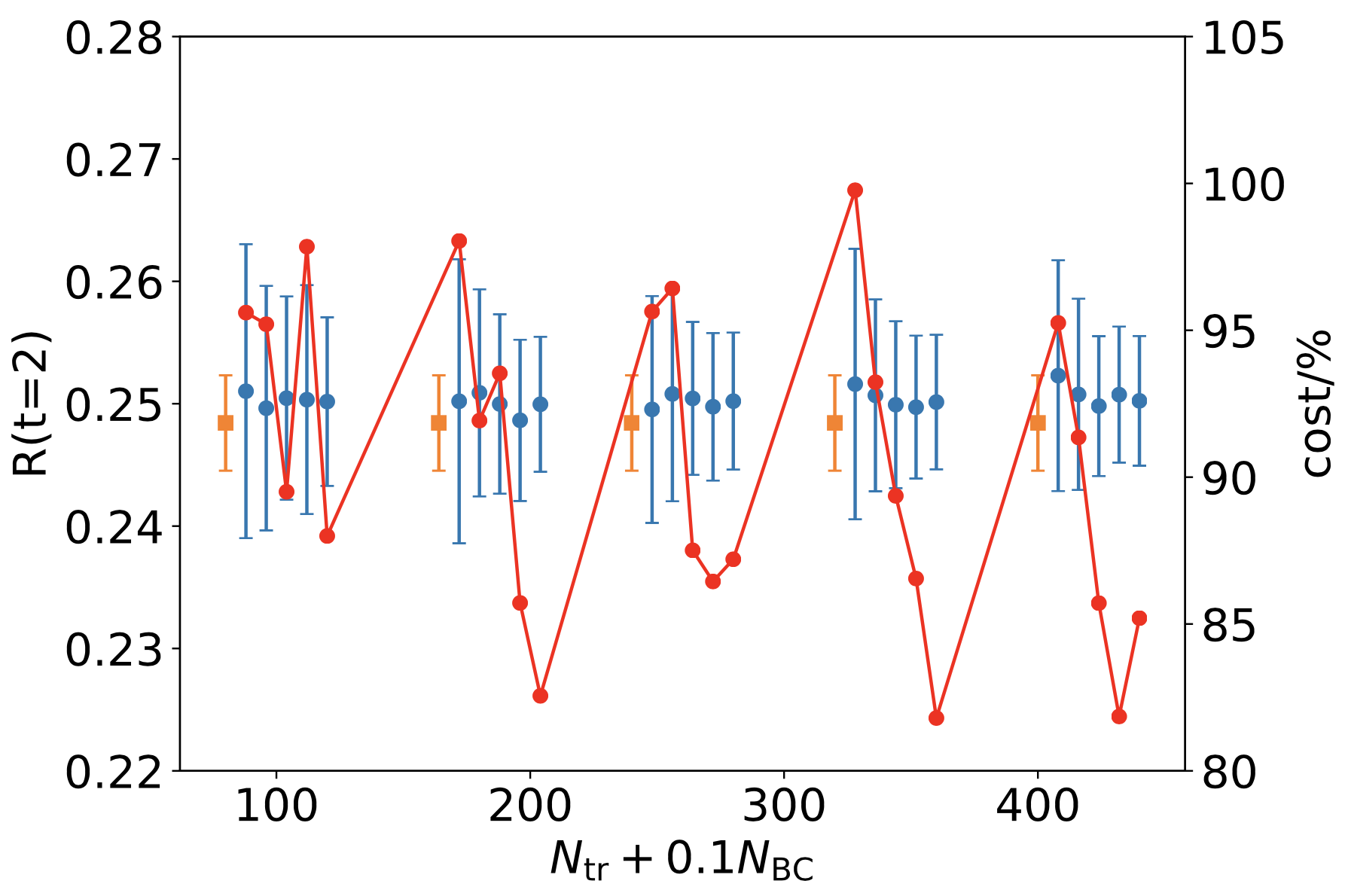}
	\caption{Predicting $R(t=2)$ (ratio of $3$-pt to $2$-pt) of kaon quasi-PDF correlators at $(p_{\mathrm{pred}},t_{\mathrm{pred}},t_{\mathrm{sep}}) = (4,4,5)$ from measurements at $(3,4,5)$. From \cite{Zhang:2019qiq}. The red curve shows the cost savings from the use of surrogate observables, with a maximum of around 20\%.\label{fig:qpdf}}
\end{figure}

It remains only to construct an approximation to the observable $\mathcal O$, and at this point the connection to machine learning is obvious. The true observable $\mathcal O$ is computed on some set of configurations, which are used to train a neural network or other device to represent the approximation $\tilde{\mathcal O}$. This approximation is by construction cheap to compute, enabling the above procedure.

This method has been applied to lattice QCD at scale, with mild success. In~\cite{Yoon:2018krb}, boosted decision trees were used to construct approximations to, among other things, nucleon three-point functions given already-computed two-point functions. That work reported reduction in computational cost of 7\%-38\%. In~\cite{Zhang:2019qiq}, both the boosted decision trees and a linear regression model were used to extract kaon quasi-PDF correlators. One figure from this latter study is reproduced in Figure~\ref{fig:qpdf}. The amount of improvement is similar to that in the previous study. A similar approach, using machine learning to extrapolate observables to longer flow times or lighter quark masses, is explored in~\cite{Kim:2024rpd}.

\section{Discussion}
We have seen four broad families of machine learning algorithms for accelerating lattice simulations: normalizing flows for improving sampling, contour deformations chiefly for improving sign problems, control variates for improving the signal-to-noise ratio, and surrogate observables for accelerating measurement. All can be used without introducing any additional systematic bias into the Monte Carlo results. Normalizing flows and surrogate observables have already been proven somewhat useful ``at scale''---at least, in $4$-dimensional gauge theories. Contour deformations and control variates have proven useful in smaller systems, but not yet in lattice QCD itself.

A recurring theme in this field is that methods that involve large numbers of parameters (usually some form of ``deep learning'') do not systematically out-perform methods with few, or even no, parameters. This was highlighted in Table~\ref{tab:flows} above with respect to normalizing flows, but the phenomenon has also repeatedly been reported in the context of contour deformations (see for example~\cite{Alexandru:2017czx,Alexandru:2018ddf}).

A second pattern, noticeable throughout this talk, is that relatively few of these methods have been successfully applied to systems of physical interest. The most remarkable exception is the study of $\mathrm{C}_{20}\mathrm{H}_{12}$-Perylene via contour deformations~\cite{Rodekamp:2024ixu}, although this work is also not quite sufficient to make direct contact with experiment.

Those interested in using these techniques---or just gaining practical experience for self-edification---may find useful a normalizing flow tutorial published several years ago~\cite{Albergo:2021vyo}, and two software packages targetting normalizing flows for lattice field theory: \cite{Tomiya:2022meu,Nicoli:2023rcd}.

\acknowledgments
This work was supported by a Richard P.~Feynman fellowship from the LANL LDRD program, and by the U.S.~Department of Energy, Office of Science, Office of High-Energy Physics under the grant KA2401045. Los Alamos National Laboratory is operated by Triad National Security, LLC, for the National Nuclear Security Administration of U.S. Department of Energy (Contract Nr. 89233218CNA000001). Report number: LA-UR-25-20975.


\bibliographystyle{JHEP}
\bibliography{ml,negahiggs}

\providecommand{\href}[2]{#2}\begingroup\raggedright\begin{thebibliography}{10}

\bibitem{Chen:2021giw}
S.Y.~Chen, H.T.~Ding, F.Y.~Liu, G.~Papp and C.B.~Yang, \emph{{Machine learning
  spectral functions in lattice QCD}},
  \href{https://arxiv.org/abs/2110.13521}{{\ttfamily 2110.13521}}.

\bibitem{SamuelOffler:2022jkv}
S.~Offler, \emph{A study of thermal NRQCD with machine learning methods},
  \href{https://doi.org/10.23889/SUthesis.60375}{Ph.D.~thesis}, Swansea U.,
  2022.

\bibitem{PhysRevLett.124.056401}
R.~Fournier, L.~Wang, O.V.~Yazyev and Q.~Wu, \emph{Artificial neural network
  approach to the analytic continuation problem},
  \href{https://doi.org/10.1103/PhysRevLett.124.056401}{\emph{Phys. Rev. Lett.}
  {\bfseries 124} (2020) 056401}.

\bibitem{PhysRevD.102.096001}
L.~Kades, J.M.~Pawlowski, A.~Rothkopf, M.~Scherzer, J.M.~Urban, S.J.~Wetzel
  et~al., \emph{Spectral reconstruction with deep neural networks},
  \href{https://doi.org/10.1103/PhysRevD.102.096001}{\emph{Phys. Rev. D}
  {\bfseries 102} (2020) 096001}.

\bibitem{Wang:2021jou}
L.~Wang, S.~Shi and K.~Zhou, \emph{{Reconstructing spectral functions via
  automatic differentiation}},
  \href{https://doi.org/10.1103/PhysRevD.106.L051502}{\emph{Phys. Rev. D}
  {\bfseries 106} (2022) L051502}
  [\href{https://arxiv.org/abs/2111.14760}{{\ttfamily 2111.14760}}].

\bibitem{carrasquilla2017machine}
J.~Carrasquilla and R.G.~Melko, \emph{Machine learning phases of matter},
  \href{https://doi.org/10.1038/nphys4035}{\emph{Nature Physics} {\bfseries 13}
  (2017) 431}.

\bibitem{Peng:2022wdl}
J.-H.~Peng, Y.-H.~Tseng and F.-J.~Jiang, \emph{{Machine learning phases of an
  Abelian gauge theory}},
  \href{https://doi.org/10.1093/ptep/ptad096}{\emph{PTEP} {\bfseries 2023}
  (2023) 073A03} [\href{https://arxiv.org/abs/2212.14655}{{\ttfamily
  2212.14655}}].

\bibitem{Tanaka:2016rtu}
A.~Tanaka and A.~Tomiya, \emph{{Detection of phase transition via convolutional
  neural network}}, \href{https://doi.org/10.7566/JPSJ.86.063001}{\emph{J.
  Phys. Soc. Jap.} {\bfseries 86} (2017) 063001}
  [\href{https://arxiv.org/abs/1609.09087}{{\ttfamily 1609.09087}}].

\bibitem{vanNieuwenburg:2016zsd}
E.P.L.~van Nieuwenburg, Y.-H.~Liu and S.D.~Huber, \emph{{Learning phase
  transitions by confusion}},
  \href{https://doi.org/10.1038/nphys4037}{\emph{Nature Phys.} {\bfseries 13}
  (2017) 435} [\href{https://arxiv.org/abs/1610.02048}{{\ttfamily
  1610.02048}}].

\bibitem{Rodriguez-Nieva:2018cbl}
J.F.~Rodriguez-Nieva and M.S.~Scheurer, \emph{{Identifying topological order
  through unsupervised machine learning}},
  \href{https://doi.org/10.1038/s41567-019-0512-x}{\emph{Nature Phys.}
  {\bfseries 15} (2019) 790}
  [\href{https://arxiv.org/abs/1805.05961}{{\ttfamily 1805.05961}}].

\bibitem{Broecker:2017hjl}
P.~Broecker, J.~Carrasquilla, R.G.~Melko and S.~Trebst, \emph{{Machine learning
  quantum phases of matter beyond the fermion sign problem}},
  \href{https://doi.org/10.1038/s41598-017-09098-0}{\emph{Sci. Rep.} {\bfseries
  7} (2017) 8823}.

\bibitem{Shanahan:2018vcv}
P.E.~Shanahan, A.~Trewartha and W.~Detmold, \emph{{Machine learning action
  parameters in lattice quantum chromodynamics}},
  \href{https://doi.org/10.1103/PhysRevD.97.094506}{\emph{Phys. Rev. D}
  {\bfseries 97} (2018) 094506}
  [\href{https://arxiv.org/abs/1801.05784}{{\ttfamily 1801.05784}}].

\bibitem{carleo2017solving}
G.~Carleo and M.~Troyer, \emph{Solving the quantum many-body problem with
  artificial neural networks},
  \href{https://doi.org/10.1126/science.aag2302}{\emph{Science} {\bfseries 355}
  (2017) 602}.

\bibitem{Deng:2017wof}
D.-L.~Deng, X.~Li and S.D.~Sarma, \emph{{Machine learning topological states}},
  \href{https://doi.org/10.1103/PhysRevB.96.195145}{\emph{Phys. Rev. B}
  {\bfseries 96} (2017) 195145}.

\bibitem{Luo:2020stn}
D.~Luo, G.~Carleo, B.K.~Clark and J.~Stokes, \emph{{Gauge Equivariant Neural
  Networks for Quantum Lattice Gauge Theories}},
  \href{https://doi.org/10.1103/PhysRevLett.127.276402}{\emph{Phys. Rev. Lett.}
  {\bfseries 127} (2021) 276402}
  [\href{https://arxiv.org/abs/2012.05232}{{\ttfamily 2012.05232}}].

\bibitem{Holland:2024muu}
K.~Holland, A.~Ipp, D.I.~M\"uller and U.~Wenger, \emph{{Machine learning a
  fixed point action for SU(3) gauge theory with a gauge equivariant
  convolutional neural network}},
  \href{https://arxiv.org/abs/2401.06481}{{\ttfamily 2401.06481}}.

\bibitem{Kanwar:2024ujc}
G.~Kanwar, \emph{Flow-based sampling for lattice field theories},
  \href{https://doi.org/10.22323/1.453.0114}{PoS \textbf{LATTICE2023} (2024)
  114} [\href{https://arxiv.org/abs/2401.01297}{2401.01297}].

\bibitem{2017PhRvB..95c5105H}
L.~{Huang} and L.~{Wang}, \emph{{Accelerated Monte Carlo simulations with
  restricted Boltzmann machines}},
  \href{https://doi.org/10.1103/PhysRevB.95.035105}{\emph{Phys. Rev. B}
  {\bfseries 95} (2017) 035105}
  [\href{https://arxiv.org/abs/1610.02746}{{\ttfamily 1610.02746}}].

\bibitem{Tanaka:2017niz}
A.~Tanaka and A.~Tomiya, \emph{{Towards reduction of autocorrelation in HMC by
  machine learning}},  \href{https://arxiv.org/abs/1712.03893}{{\ttfamily
  1712.03893}}.

\bibitem{Wang:2023exq}
L.~Wang, G.~Aarts and K.~Zhou, \emph{{Diffusion models as stochastic
  quantization in lattice field theory}},
  \href{https://doi.org/10.1007/JHEP05(2024)060}{\emph{JHEP} {\bfseries 05}
  (2024) 060} [\href{https://arxiv.org/abs/2309.17082}{{\ttfamily
  2309.17082}}].

\bibitem{Wang:2023sry}
L.~Wang, G.~Aarts and K.~Zhou, \emph{{Generative Diffusion Models for Lattice
  Field Theory}},  in \emph{{37th Conference on Neural Information Processing
  Systems}}, 11, 2023 \href{https://arxiv.org/abs/2311.03578}{{\ttfamily
  2311.03578}}.

\bibitem{Zhou:2018ill}
K.~Zhou, G.~Endr\H{o}di, L.-G.~Pang and H.~St\"ocker, \emph{{Regressive and
  generative neural networks for scalar field theory}},
  \href{https://doi.org/10.1103/PhysRevD.100.011501}{\emph{Phys. Rev. D}
  {\bfseries 100} (2019) 011501}
  [\href{https://arxiv.org/abs/1810.12879}{{\ttfamily 1810.12879}}].

\bibitem{Pawlowski:2018qxs}
J.M.~Pawlowski and J.M.~Urban, \emph{{Reducing Autocorrelation Times in Lattice
  Simulations with Generative Adversarial Networks}},
  \href{https://doi.org/10.1088/2632-2153/abae73}{\emph{Mach. Learn. Sci.
  Tech.} {\bfseries 1} (2020) 045011}
  [\href{https://arxiv.org/abs/1811.03533}{{\ttfamily 1811.03533}}].

\bibitem{box1958note}
G.E.~Box and M.E.~Muller, \emph{A note on the generation of random normal
  deviates}, {\emph{The annals of mathematical statistics} {\bfseries 29}
  (1958) 610}.

\bibitem{Albergo:2019eim}
M.S.~Albergo, G.~Kanwar and P.E.~Shanahan, \emph{{Flow-based generative models
  for Markov chain Monte Carlo in lattice field theory}},
  \href{https://doi.org/10.1103/PhysRevD.100.034515}{\emph{Phys. Rev. D}
  {\bfseries 100} (2019) 034515}
  [\href{https://arxiv.org/abs/1904.12072}{{\ttfamily 1904.12072}}].

\bibitem{Nicolai:1979nr}
H.~Nicolai, \emph{{On a New Characterization of Scalar Supersymmetric
  Theories}}, \href{https://doi.org/10.1016/0370-2693(80)90138-0}{\emph{Phys.
  Lett. B} {\bfseries 89} (1980) 341}.

\bibitem{Luscher:2009eq}
M.~Luscher, \emph{{Trivializing maps, the Wilson flow and the HMC algorithm}},
  \href{https://doi.org/10.1007/s00220-009-0953-7}{\emph{Commun. Math. Phys.}
  {\bfseries 293} (2010) 899}
  [\href{https://arxiv.org/abs/0907.5491}{{\ttfamily 0907.5491}}].

\bibitem{2014arXiv1410.8516D}
L.~{Dinh}, D.~{Krueger} and Y.~{Bengio}, \emph{{NICE: Non-linear Independent
  Components Estimation}},  \href{https://arxiv.org/abs/1410.8516}{{\ttfamily
  1410.8516}}.

\bibitem{2016arXiv160508803D}
L.~{Dinh}, J.~{Sohl-Dickstein} and S.~{Bengio}, \emph{{Density estimation using
  Real NVP}},  \href{https://arxiv.org/abs/1605.08803}{{\ttfamily 1605.08803}}.

\bibitem{2018arXiv180703039K}
D.P.~{Kingma} and P.~{Dhariwal}, \emph{{Glow: Generative Flow with Invertible
  1x1 Convolutions}},  \href{https://arxiv.org/abs/1807.03039}{{\ttfamily
  1807.03039}}.

\bibitem{Caselle:2023mvh}
M.~Caselle, E.~Cellini and A.~Nada, \emph{{Sampling the lattice Nambu-Goto
  string using Continuous Normalizing Flows}},
  \href{https://doi.org/10.1007/JHEP02(2024)048}{\emph{JHEP} {\bfseries 02}
  (2024) 048} [\href{https://arxiv.org/abs/2307.01107}{{\ttfamily
  2307.01107}}].

\bibitem{deHaan:2021erb}
P.~de~Haan, C.~Rainone, M.C.N.~Cheng and R.~Bondesan, \emph{{Scaling Up Machine
  Learning For Quantum Field Theory with Equivariant Continuous Flows}},
  \href{https://arxiv.org/abs/2110.02673}{{\ttfamily 2110.02673}}.

\bibitem{Nicoli:2019gun}
K.A.~Nicoli, S.~Nakajima, N.~Strodthoff, W.~Samek, K.-R.~M\"uller and
  P.~Kessel, \emph{{Asymptotically unbiased estimation of physical observables
  with neural samplers}},
  \href{https://doi.org/10.1103/PhysRevE.101.023304}{\emph{Phys. Rev. E}
  {\bfseries 101} (2020) 023304}
  [\href{https://arxiv.org/abs/1910.13496}{{\ttfamily 1910.13496}}].

\bibitem{Nicoli:2020njz}
K.A.~Nicoli, C.J.~Anders, L.~Funcke, T.~Hartung, K.~Jansen, P.~Kessel et~al.,
  \emph{{Estimation of Thermodynamic Observables in Lattice Field Theories with
  Deep Generative Models}},
  \href{https://doi.org/10.1103/PhysRevLett.126.032001}{\emph{Phys. Rev. Lett.}
  {\bfseries 126} (2021) 032001}
  [\href{https://arxiv.org/abs/2007.07115}{{\ttfamily 2007.07115}}].

\bibitem{Boyda:2020hsi}
D.~Boyda, G.~Kanwar, S.~Racani\`ere, D.J.~Rezende, M.S.~Albergo, K.~Cranmer
  et~al., \emph{{Sampling using $SU(N)$ gauge equivariant flows}},
  \href{https://doi.org/10.1103/PhysRevD.103.074504}{\emph{Phys. Rev. D}
  {\bfseries 103} (2021) 074504}
  [\href{https://arxiv.org/abs/2008.05456}{{\ttfamily 2008.05456}}].

\bibitem{Kanwar:2020xzo}
G.~Kanwar, M.S.~Albergo, D.~Boyda, K.~Cranmer, D.C.~Hackett, S.~Racani\`ere
  et~al., \emph{{Equivariant flow-based sampling for lattice gauge theory}},
  \href{https://doi.org/10.1103/PhysRevLett.125.121601}{\emph{Phys. Rev. Lett.}
  {\bfseries 125} (2020) 121601}
  [\href{https://arxiv.org/abs/2003.06413}{{\ttfamily 2003.06413}}].

\bibitem{Kanwar:2021wzm}
G.~Kanwar, \emph{{Machine Learning and Variational Algorithms for Lattice Field
  Theory}}, Ph.D. thesis, MIT, 2021.
\newblock \href{https://arxiv.org/abs/2106.01975}{{\ttfamily 2106.01975}}.

\bibitem{Abbott:2022zhs}
R.~Abbott et~al., \emph{{Gauge-equivariant flow models for sampling in lattice
  field theories with pseudofermions}},
  \href{https://doi.org/10.1103/PhysRevD.106.074506}{\emph{Phys. Rev. D}
  {\bfseries 106} (2022) 074506}
  [\href{https://arxiv.org/abs/2207.08945}{{\ttfamily 2207.08945}}].

\bibitem{Albergo:2022qfi}
M.S.~Albergo, D.~Boyda, K.~Cranmer, D.C.~Hackett, G.~Kanwar, S.~Racani\`ere
  et~al., \emph{{Flow-based sampling in the lattice Schwinger model at
  criticality}}, \href{https://doi.org/10.1103/PhysRevD.106.014514}{\emph{Phys.
  Rev. D} {\bfseries 106} (2022) 014514}
  [\href{https://arxiv.org/abs/2202.11712}{{\ttfamily 2202.11712}}].

\bibitem{Abbott:2023thq}
R.~Abbott et~al., \emph{{Normalizing flows for lattice gauge theory in
  arbitrary space-time dimension}},
  \href{https://arxiv.org/abs/2305.02402}{{\ttfamily 2305.02402}}.

\bibitem{Bacchio:2022vje}
S.~Bacchio, P.~Kessel, S.~Schaefer and L.~Vaitl, \emph{{Learning trivializing
  gradient flows for lattice gauge theories}},
  \href{https://doi.org/10.1103/PhysRevD.107.L051504}{\emph{Phys. Rev. D}
  {\bfseries 107} (2023) L051504}
  [\href{https://arxiv.org/abs/2212.08469}{{\ttfamily 2212.08469}}].

\bibitem{Lawrence:2022afv}
S.~Lawrence, H.~Oh and Y.~Yamauchi, \emph{{Lattice scalar field theory at
  complex coupling}},
  \href{https://doi.org/10.1103/PhysRevD.106.114503}{\emph{Phys. Rev. D}
  {\bfseries 106} (2022) 114503}
  [\href{https://arxiv.org/abs/2205.12303}{{\ttfamily 2205.12303}}].

\bibitem{Rodekamp:2024ixu}
M.~Rodekamp, E.~Berkowitz, C.~G\"antgen, S.~Krieg, T.~Luu, J.~Ostmeyer et~al.,
  \emph{{Single Particle Spectrum of Doped
  $\mathrm{C}_{20}\mathrm{H}_{12}$-Perylene}},
  \href{https://arxiv.org/abs/2406.06711}{{\ttfamily 2406.06711}}.

\bibitem{Alexandru:2020wrj}
A.~Alexandru, G.~Basar, P.F.~Bedaque and N.C.~Warrington, \emph{{Complex paths
  around the sign problem}},
  \href{https://doi.org/10.1103/RevModPhys.94.015006}{\emph{Rev. Mod. Phys.}
  {\bfseries 94} (2022) 015006}
  [\href{https://arxiv.org/abs/2007.05436}{{\ttfamily 2007.05436}}].

\bibitem{Witten:2010cx}
E.~Witten, \emph{{Analytic Continuation Of Chern-Simons Theory}}, {\emph{AMS/IP
  Stud. Adv. Math.} {\bfseries 50} (2011) 347}
  [\href{https://arxiv.org/abs/1001.2933}{{\ttfamily 1001.2933}}].

\bibitem{Cristoforetti:2012su}
{\scshape AuroraScience} collaboration, \emph{{New approach to the sign problem
  in quantum field theories: High density QCD on a Lefschetz thimble}},
  \href{https://doi.org/10.1103/PhysRevD.86.074506}{\emph{Phys. Rev. D}
  {\bfseries 86} (2012) 074506}
  [\href{https://arxiv.org/abs/1205.3996}{{\ttfamily 1205.3996}}].

\bibitem{Alexandru:2017czx}
A.~Alexandru, P.F.~Bedaque, H.~Lamm and S.~Lawrence, \emph{{Deep Learning
  Beyond Lefschetz Thimbles}},
  \href{https://doi.org/10.1103/PhysRevD.96.094505}{\emph{Phys. Rev. D}
  {\bfseries 96} (2017) 094505}
  [\href{https://arxiv.org/abs/1709.01971}{{\ttfamily 1709.01971}}].

\bibitem{Mori:2017pne}
Y.~Mori, K.~Kashiwa and A.~Ohnishi, \emph{{Toward solving the sign problem with
  path optimization method}},
  \href{https://doi.org/10.1103/PhysRevD.96.111501}{\emph{Phys. Rev. D}
  {\bfseries 96} (2017) 111501}
  [\href{https://arxiv.org/abs/1705.05605}{{\ttfamily 1705.05605}}].

\bibitem{Alexandru:2018fqp}
A.~Alexandru, P.F.~Bedaque, H.~Lamm and S.~Lawrence, \emph{{Finite-Density
  Monte Carlo Calculations on Sign-Optimized Manifolds}},
  \href{https://doi.org/10.1103/PhysRevD.97.094510}{\emph{Phys. Rev. D}
  {\bfseries 97} (2018) 094510}
  [\href{https://arxiv.org/abs/1804.00697}{{\ttfamily 1804.00697}}].

\bibitem{Alexandru:2018ddf}
A.~Alexandru, P.F.~Bedaque, H.~Lamm, S.~Lawrence and N.C.~Warrington,
  \emph{{Fermions at Finite Density in 2+1 Dimensions with Sign-Optimized
  Manifolds}},
  \href{https://doi.org/10.1103/PhysRevLett.121.191602}{\emph{Phys. Rev. Lett.}
  {\bfseries 121} (2018) 191602}
  [\href{https://arxiv.org/abs/1808.09799}{{\ttfamily 1808.09799}}].

\bibitem{Gantgen:2023byf}
C.~G\"antgen, E.~Berkowitz, T.~Luu, J.~Ostmeyer and M.~Rodekamp,
  \emph{{Fermionic sign problem minimization by constant path integral contour
  shifts}}, \href{https://doi.org/10.1103/PhysRevB.109.195158}{\emph{Phys. Rev.
  B} {\bfseries 109} (2024) 195158}
  [\href{https://arxiv.org/abs/2307.06785}{{\ttfamily 2307.06785}}].

\bibitem{Bender:2018pbv}
C.M.~Bender, N.~Hassanpour, S.P.~Klevansky and S.~Sarkar, \emph{{$PT$-symmetric
  quantum field theory in $D$ dimensions}},
  \href{https://doi.org/10.1103/PhysRevD.98.125003}{\emph{Phys. Rev. D}
  {\bfseries 98} (2018) 125003}
  [\href{https://arxiv.org/abs/1810.12479}{{\ttfamily 1810.12479}}].

\bibitem{Romatschke:2024cld}
P.~Romatschke, \emph{{On the negative coupling O(N) model in 2d at high
  temperature}},  \href{https://arxiv.org/abs/2412.10496}{{\ttfamily
  2412.10496}}.

\bibitem{Lawrence:2023woz}
S.~Lawrence, R.~Weller, C.~Peterson and P.~Romatschke, \emph{{Instantons,
  analytic continuation, and PT-symmetric field theory}},
  \href{https://doi.org/10.1103/PhysRevD.108.085013}{\emph{Phys. Rev. D}
  {\bfseries 108} (2023) 085013}
  [\href{https://arxiv.org/abs/2303.01470}{{\ttfamily 2303.01470}}].

\bibitem{Weller:2023jhc}
R.D.~Weller, \emph{{Can negative bare couplings make sense? The $\vec{\phi}^4$
  theory at large $N$}},  \href{https://arxiv.org/abs/2310.02516}{{\ttfamily
  2310.02516}}.

\bibitem{Tulipant:2022vtk}
Z.~Tulipant, M.~Giordano, K.~Kapas, S.D.~Katz and A.~Pasztor,
  \emph{{Exponential improvement of the sign problem via contour deformations
  in the 2+1D XY model at non-zero density}},
  \href{https://doi.org/10.22323/1.430.0161}{\emph{Phys. Rev. D} {\bfseries
  106} (2022) 054512} [\href{https://arxiv.org/abs/2202.07561}{{\ttfamily
  2202.07561}}].

\bibitem{Kashiwa:2020brj}
K.~Kashiwa and Y.~Mori, \emph{{Path optimization for $U(1)$ gauge theory with
  complexified parameters}},
  \href{https://doi.org/10.1103/PhysRevD.102.054519}{\emph{Phys. Rev. D}
  {\bfseries 102} (2020) 054519}
  [\href{https://arxiv.org/abs/2007.04167}{{\ttfamily 2007.04167}}].

\bibitem{Basar:2022cef}
G.~Basar and J.~Marincel, \emph{{Sign optimization and complex saddle points in
  one-dimensional QCD}},
  \href{https://doi.org/10.1103/PhysRevD.106.L091503}{\emph{Phys. Rev. D}
  {\bfseries 106} (2022) L091503}
  [\href{https://arxiv.org/abs/2208.02072}{{\ttfamily 2208.02072}}].

\bibitem{Basar:2023bwd}
G.~Basar and J.~Marincel, \emph{{Heavy-dense QCD, sign optimization, and
  Lefschetz thimbles}},
  \href{https://doi.org/10.1103/PhysRevC.109.045208}{\emph{Phys. Rev. C}
  {\bfseries 109} (2024) 045208}
  [\href{https://arxiv.org/abs/2311.06343}{{\ttfamily 2311.06343}}].

\bibitem{Namekawa:2021nzu}
Y.~Namekawa, K.~Kashiwa, A.~Ohnishi and H.~Takase, \emph{{Gauge invariant input
  to neural network for path optimization method}},
  \href{https://doi.org/10.1103/PhysRevD.105.034502}{\emph{Phys. Rev. D}
  {\bfseries 105} (2022) 034502}
  [\href{https://arxiv.org/abs/2109.11710}{{\ttfamily 2109.11710}}].

\bibitem{Giordano:2023ppk}
M.~Giordano, A.~Pasztor, D.~Pesznyak and Z.~Tulipant, \emph{{Alleviating the
  sign problem in a chiral random matrix model with contour deformations}},
  \href{https://doi.org/10.1103/PhysRevD.108.094507}{\emph{Phys. Rev. D}
  {\bfseries 108} (2023) 094507}
  [\href{https://arxiv.org/abs/2301.12947}{{\ttfamily 2301.12947}}].

\bibitem{Alexandru:2018ngw}
A.~Alexandru, G.~Ba\c{s}ar, P.F.~Bedaque, H.~Lamm and S.~Lawrence,
  \emph{{Finite Density $QED_{1+1}$ Near Lefschetz Thimbles}},
  \href{https://doi.org/10.1103/PhysRevD.98.034506}{\emph{Phys. Rev. D}
  {\bfseries 98} (2018) 034506}
  [\href{https://arxiv.org/abs/1807.02027}{{\ttfamily 1807.02027}}].

\bibitem{Kashiwa:2023dfx}
K.~Kashiwa, Y.~Namekawa, A.~Ohnishi and H.~Takase, \emph{{Application of the
  path optimization method to a discrete spin system}},
  \href{https://doi.org/10.1103/PhysRevD.108.094504}{\emph{Phys. Rev. D}
  {\bfseries 108} (2023) 094504}
  [\href{https://arxiv.org/abs/2309.06018}{{\ttfamily 2309.06018}}].

\bibitem{Warrington:2023aqa}
N.C.~Warrington, \emph{{Real-time spin systems from lattice field theory}},
  \href{https://doi.org/10.1007/JHEP12(2023)156}{\emph{JHEP} {\bfseries 12}
  (2023) 156} [\href{https://arxiv.org/abs/2310.19761}{{\ttfamily
  2310.19761}}].

\bibitem{Mooney:2021esz}
T.C.~Mooney, J.~Bringewatt, N.C.~Warrington and L.T.~Brady, \emph{{Lefschetz
  thimble quantum Monte Carlo for spin systems}},
  \href{https://doi.org/10.1103/PhysRevB.106.214416}{\emph{Phys. Rev. B}
  {\bfseries 106} (2022) 214416}
  [\href{https://arxiv.org/abs/2110.10699}{{\ttfamily 2110.10699}}].

\bibitem{Alexandru:2017lqr}
A.~Alexandru, G.~Basar, P.F.~Bedaque and G.W.~Ridgway, \emph{{Schwinger-Keldysh
  formalism on the lattice: A faster algorithm and its application to field
  theory}}, \href{https://doi.org/10.1103/PhysRevD.95.114501}{\emph{Phys. Rev.
  D} {\bfseries 95} (2017) 114501}
  [\href{https://arxiv.org/abs/1704.06404}{{\ttfamily 1704.06404}}].

\bibitem{Alexandru:2016gsd}
A.~Alexandru, G.~Basar, P.F.~Bedaque, S.~Vartak and N.C.~Warrington,
  \emph{{Monte Carlo Study of Real Time Dynamics on the Lattice}},
  \href{https://doi.org/10.1103/PhysRevLett.117.081602}{\emph{Phys. Rev. Lett.}
  {\bfseries 117} (2016) 081602}
  [\href{https://arxiv.org/abs/1605.08040}{{\ttfamily 1605.08040}}].

\bibitem{Lawrence:2021izu}
S.~Lawrence and Y.~Yamauchi, \emph{{Normalizing Flows and the Real-Time Sign
  Problem}}, \href{https://doi.org/10.1103/PhysRevD.103.114509}{\emph{Phys.
  Rev. D} {\bfseries 103} (2021) 114509}
  [\href{https://arxiv.org/abs/2101.05755}{{\ttfamily 2101.05755}}].

\bibitem{Kanwar:2021tkd}
G.~Kanwar and M.L.~Wagman, \emph{{Real-time lattice gauge theory actions:
  Unitarity, convergence, and path integral contour deformations}},
  \href{https://doi.org/10.1103/PhysRevD.104.014513}{\emph{Phys. Rev. D}
  {\bfseries 104} (2021) 014513}
  [\href{https://arxiv.org/abs/2103.02602}{{\ttfamily 2103.02602}}].

\bibitem{Detmold:2021ulb}
W.~Detmold, G.~Kanwar, H.~Lamm, M.L.~Wagman and N.C.~Warrington, \emph{{Path
  integral contour deformations for observables in $SU(N)$ gauge theory}},
  \href{https://doi.org/10.1103/PhysRevD.103.094517}{\emph{Phys. Rev. D}
  {\bfseries 103} (2021) 094517}
  [\href{https://arxiv.org/abs/2101.12668}{{\ttfamily 2101.12668}}].

\bibitem{Detmold:2020ncp}
W.~Detmold, G.~Kanwar, M.L.~Wagman and N.C.~Warrington, \emph{{Path integral
  contour deformations for noisy observables}},
  \href{https://doi.org/10.1103/PhysRevD.102.014514}{\emph{Phys. Rev. D}
  {\bfseries 102} (2020) 014514}
  [\href{https://arxiv.org/abs/2003.05914}{{\ttfamily 2003.05914}}].

\bibitem{Kanwar:2023otc}
G.~Kanwar, A.~Lovato, N.~Rocco and M.~Wagman, \emph{{Mitigating Green's
  function Monte Carlo signal-to-noise problems using contour deformations}},
  \href{https://doi.org/10.1103/PhysRevC.109.034317}{\emph{Phys. Rev. C}
  {\bfseries 109} (2024) 034317}
  [\href{https://arxiv.org/abs/2304.03229}{{\ttfamily 2304.03229}}].

\bibitem{Lawrence:2024pjg}
S.~Lawrence, S.~Valgushev, J.~Xiao and Y.~Yamauchi, \emph{{Contour deformations
  for non-holomorphic actions}},
  \href{https://arxiv.org/abs/2401.16733}{{\ttfamily 2401.16733}}.

\bibitem{Lawrence:2023sfc}
S.~Lawrence and Y.~Yamauchi, \emph{{Convex optimization of contour
  deformations}},
  \href{https://doi.org/10.1103/PhysRevD.110.014508}{\emph{Phys. Rev. D}
  {\bfseries 110} (2024) 014508}
  [\href{https://arxiv.org/abs/2311.13002}{{\ttfamily 2311.13002}}].

\bibitem{Abbott:2024kfc}
R.~Abbott, A.~Botev, D.~Boyda, D.C.~Hackett, G.~Kanwar, S.~Racani\`ere et~al.,
  \emph{{Applications of flow models to the generation of correlated lattice
  QCD ensembles}},
  \href{https://doi.org/10.1103/PhysRevD.109.094514}{\emph{Phys. Rev. D}
  {\bfseries 109} (2024) 094514}
  [\href{https://arxiv.org/abs/2401.10874}{{\ttfamily 2401.10874}}].

\bibitem{Lawrence:2022dba}
S.~Lawrence and Y.~Yamauchi, \emph{{Deep learning of fermion sign
  fluctuations}},
  \href{https://doi.org/10.1103/PhysRevD.107.114505}{\emph{Phys. Rev. D}
  {\bfseries 107} (2023) 114505}
  [\href{https://arxiv.org/abs/2212.14606}{{\ttfamily 2212.14606}}].

\bibitem{Lawrence:2020kyw}
S.~Lawrence, \emph{{Perturbative Removal of a Sign Problem}},
  \href{https://doi.org/10.1103/PhysRevD.102.094504}{\emph{Phys. Rev. D}
  {\bfseries 102} (2020) 094504}
  [\href{https://arxiv.org/abs/2009.10901}{{\ttfamily 2009.10901}}].

\bibitem{Lawrence:2023cft}
S.~Lawrence and Y.~Yamauchi, \emph{{Mitigating a discrete sign problem with
  extreme learning machines}},
  \href{https://arxiv.org/abs/2312.12636}{{\ttfamily 2312.12636}}.

\bibitem{Bhattacharya:2023pxx}
T.~Bhattacharya, S.~Lawrence and J.-S.~Yoo, \emph{{Control variates for lattice
  field theory}},
  \href{https://doi.org/10.1103/PhysRevD.109.L031505}{\emph{Phys. Rev. D}
  {\bfseries 109} (2024) L031505}
  [\href{https://arxiv.org/abs/2307.14950}{{\ttfamily 2307.14950}}].

\bibitem{Bedaque:2023ovz}
P.F.~Bedaque and H.~Oh, \emph{{Leveraging neural control variates for enhanced
  precision in lattice field theory}},
  \href{https://doi.org/10.1103/PhysRevD.109.094519}{\emph{Phys. Rev. D}
  {\bfseries 109} (2024) 094519}
  [\href{https://arxiv.org/abs/2312.08228}{{\ttfamily 2312.08228}}].

\bibitem{Lawrence:2024xsi}
S.~Lawrence, \emph{{Schwinger-Dyson control variates for lattice fermions}},
  \href{https://arxiv.org/abs/2404.10707}{{\ttfamily 2404.10707}}.

\bibitem{Zhang:2019qiq}
R.~Zhang, Z.~Fan, R.~Li, H.-W.~Lin and B.~Yoon, \emph{{Machine-learning
  prediction for quasiparton distribution function matrix elements}},
  \href{https://doi.org/10.1103/PhysRevD.101.034516}{\emph{Phys. Rev. D}
  {\bfseries 101} (2020) 034516}
  [\href{https://arxiv.org/abs/1909.10990}{{\ttfamily 1909.10990}}].

\bibitem{Yoon:2018krb}
B.~Yoon, T.~Bhattacharya and R.~Gupta, \emph{{Machine Learning Estimators for
  Lattice QCD Observables}},
  \href{https://doi.org/10.1103/PhysRevD.100.014504}{\emph{Phys. Rev. D}
  {\bfseries 100} (2019) 014504}
  [\href{https://arxiv.org/abs/1807.05971}{{\ttfamily 1807.05971}}].

\bibitem{Kim:2024rpd}
J.~Kim, G.~Pederiva and A.~Shindler, \emph{{Machine learning mapping of lattice
  correlated data}},
  \href{https://doi.org/10.1016/j.physletb.2024.138894}{\emph{Phys. Lett. B}
  {\bfseries 856} (2024) 138894}
  [\href{https://arxiv.org/abs/2402.07450}{{\ttfamily 2402.07450}}].

\bibitem{Albergo:2021vyo}
M.S.~Albergo, D.~Boyda, D.C.~Hackett, G.~Kanwar, K.~Cranmer, S.~Racani\`ere
  et~al., \emph{{Introduction to Normalizing Flows for Lattice Field Theory}},
  \href{https://arxiv.org/abs/2101.08176}{{\ttfamily 2101.08176}}.

\bibitem{Tomiya:2022meu}
A.~Tomiya and S.~Terasaki, \emph{{GomalizingFlow.jl: A Julia package for
  Flow-based sampling algorithm for lattice field theory}},
  \href{https://arxiv.org/abs/2208.08903}{{\ttfamily 2208.08903}}.

\bibitem{Nicoli:2023rcd}
K.A.~Nicoli, C.J.~Anders, L.~Funcke, K.~Jansen, S.~Nakajima and P.~Kessel,
  \emph{{NeuLat: a toolbox for neural samplers in lattice field theories}},
  \href{https://doi.org/10.22323/1.453.0286}{\emph{PoS} {\bfseries LATTICE2023}
  (2024) 286}.

\end{thebibliography}\endgroup

\end{document}